\documentclass{article}

\usepackage{arxiv}

\usepackage[utf8]{inputenc} 
\usepackage[T1]{fontenc}    
\usepackage{hyperref}       
\usepackage{url}            
\usepackage{booktabs}       
\usepackage{amsfonts,amsmath,amssymb}       
\usepackage{nicefrac}       
\usepackage{microtype}      
\usepackage{graphicx}
\usepackage{caption}

\usepackage{color, colortbl}

\usepackage{mathptmx,amsthm}
\usepackage{xcolor}
\usepackage{multirow}
\usepackage{subcaption}

\newcommand{\etal}{\textit{et al.}}
\newcommand{\ie}{\textit{i.e.,}}
\newcommand{\eg}{\textit{e.g.,}}
\newcommand{\commenttxt}[1]{}

\newcommand{\mybar}{\kern1pt\rule[-\dp\strutbox]{.8pt}{\baselineskip}\kern1pt}

\renewcommand{\mytagline}{} 

\theoremstyle{definition}

\title{Adversarial Robustness of Deep Learning Models for Inland Water Body Segmentation from SAR Images}
\author{
  Siddharth Kothari\textsuperscript{1},
  Srinivasan Murali\textsuperscript{1},
  Sankalp Kothari\textsuperscript{1},
  Ujjwal Verma\textsuperscript{2}\thanks{\texttt{ujjwal.verma@manipal.edu}},
  Jaya~Sreevalsan-Nair\textsuperscript{1}\thanks{\texttt{jnair@iiitb.ac.in}} 
  \\
  \textsuperscript{1}Graphics-Visualization-Computing Lab,\\
  International Institute of Information Technology Bangalore, Karnataka 560100, India. \\
  \texttt{http://www.iiitb.ac.in/gvcl} \\
  \textsuperscript{2}Department of Electronics and Communication Engineering, \\
  Manipal Institute of Technology, Manipal Academy of Higher Education, Manipal, Karnataka 576104, India.
}

\begin{document}
\maketitle

\begin{abstract}
Inland water body segmentation from Synthetic Aperture Radar (SAR) images is an important task needed for several applications, such as flood mapping. While SAR sensors capture data in all-weather conditions as high-resolution images, differentiating water and water-like surfaces from SAR images is not straightforward. Inland water bodies, such as large river basins, have complex geometry, which adds to the challenge of segmentation. U-Net is a widely used deep learning model for land-water segmentation of SAR images. In practice, manual annotation is often used to generate the corresponding water masks as ground truth. Manual annotation of the images is prone to label noise owing to data poisoning attacks, especially due to complex geometry. In this work, we simulate manual errors in the form of adversarial attacks on the U-Net model and study the robustness of the model to human errors in annotation. Our results indicate that U-Net can tolerate a certain level of corruption before its performance drops significantly. This finding highlights the crucial role that the quality of manual annotations plays in determining the effectiveness of the segmentation model. The code and the new dataset, along with adversarial examples for robust training, are publicly available.
(GitHub link - \url{https://github.com/GVCL/IWSeg-SAR-Poison.git})
\end{abstract}


\section{Introduction}
\begin{figure}
\centering
\includegraphics[width=\columnwidth]{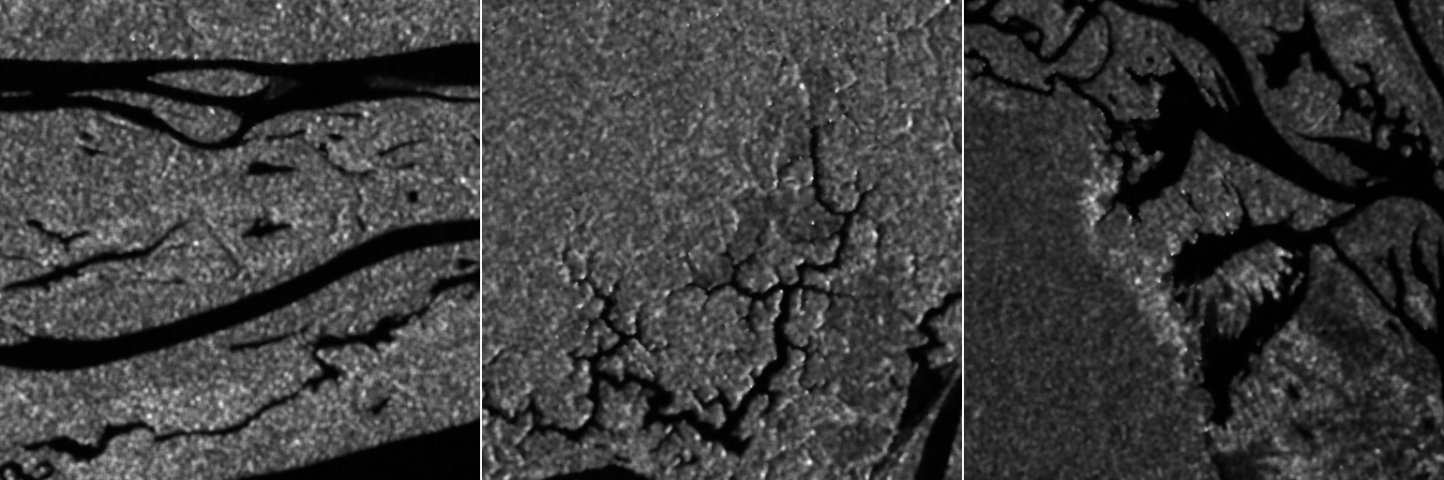}
\caption{Samples of SAR images from Sentinel-1 with inland water bodies with complex geometry in the Amazon River Basin, which is a region of interest in our work.}
\label{fig:samples}
\end{figure}

Water body segmentation is a critical task in remote sensing and geospatial analysis, which caters to diverse and critical applications such as flood mapping and monitoring~\cite{8258373}, glacial lake segmentation~\cite{WANG2022289}, sedimentation and erosion studies~\cite{OSULLIVAN2024101276}, and environmental impact assessments. These tasks often rely on high-resolution satellite imagery, such as Synthetic Aperture Radar (SAR), and consequently, accurate segmentation models to delineate water bodies from surrounding land features. SAR imaging uses energy reflectance to be available through day-night lighting and all-weather conditions. However, SAR sensors do not capture data that can discriminate between water and water-like surfaces~\cite{pech2023sentinel}. Inland water body segmentation is far more challenging than coastline detection, when discussing land-water segmentation of SAR images, owing to the complex geometry of hydrological structures, such as river basins with several tributaries and distributaries (Figure~\ref{fig:samples}).

Deep learning (DL) models, namely, convolutional neural networks (CNNs), outperform traditional approaches in SAR image segmentation in several scenarios. Particularly, encoder-decoder architectures in CNNs, such as U-Net~\cite{ronneberger2015u}, have demonstrated state-of-the-art performance in land-water segmentation of SAR images. Despite these advances in DL, water body segmentation models continue to face challenges during the learning process, such as the presence of noise, domain shifts in remote sensing data, and the impact of adversarial perturbations. Several works in this area rely primarily on manual annotation to generate ground truth masks~\cite{8258373,erfani2022atlantis,xia2025openearthmapsarbenchmarksyntheticaperture}. These are prone to several human errors, especially during manual annotation. These inadvertent errors include incorrect labels due to variations in river width and mislabeling boundary pixels. These errors tend to propagate to the image segmentation models using DL, which calls for testing the robustness of these architectures. This motivates our investigation into potential labeling errors and their simulation. Our approach in this study is the novel usage of \textit{morphological noise}, followed by a systematic testing of adversarial learning of segmentation models, specifically the U-Net architecture. Morphological noise closely models the incorrectness in labels owing to the complex geometry of the hydrological structures of a large river basin, \eg~the Amazon River Basin in South America.

Adversarial attacks pose a significant threat to the robustness and reliability of machine learning models. These attacks involve intentionally crafted perturbations to input data that cause models to misclassify or produce incorrect outputs, often with changes that are barely discernible to the human eye. In the context of water body segmentation, adversarial attacks can lead to inaccurate delineation of water boundaries, with potentially severe consequences for downstream applications such as flood risk assessment or glacial melt monitoring~\cite{Rahaman18082022,isprs-archives-XLVIII-M-3-2023-81-2023}. In this work, the primary form of corruption that we experiment with is \textit{data poisoning attacks}, which involve corruption of the training data passed to the model. We introduce a specific type of noise, which is the \textit{morphological label noise}, that is generated using morphological operations (erosion and dilation) in a controlled manner to introduce corruptions in the data points, thus simulating boundary errors.

A recent work has highlighted the vulnerability of segmentation models to adversarial attacks, including both pixel-level perturbations and structure-aware attacks that exploit model vulnerabilities~\cite{gu2021adversarialexamplessegmentationmodels}. While adversarial robustness has been extensively studied in tasks like image classification~\cite{dong2020benchmarking}, its impact on geospatial tasks, particularly water body segmentation, remains underexplored. This is especially of interest to the community given the high stakes associated with decisions informed by segmentation outputs in environmental and disaster management contexts. There is recent work which highlights the importance of training with noisy labels to improve robustness of the models~\cite{liu2024task}.

\begin{figure*}[htbp]
    \centering
    \includegraphics[width=\linewidth]{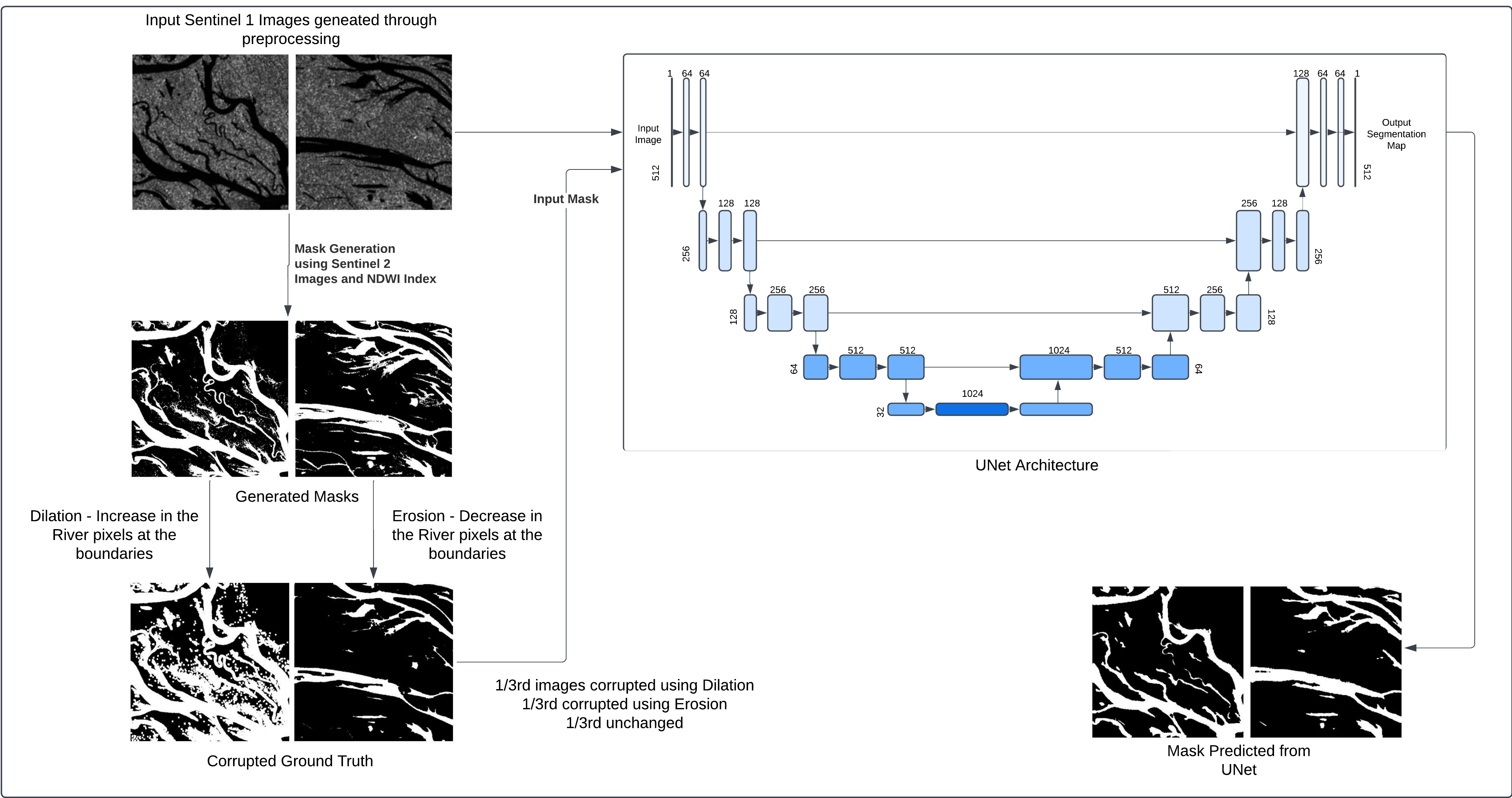}
    \caption{The workflow of our proposed method used for simulating \textit{adversarial attacks} in ground truth generation for inland water body segmentation.}
    \label{fig:flowchart}
\end{figure*}

In this paper, we investigate the influence of adversarial attacks in training inland water body segmentation models. The primary type of attacks that we focus on are data poisoning attacks through corruption of the training data samples, thus leading to prediction errors. Specifically, we explore the vulnerabilities of U-Net-based models under these settings. Our contributions include:

\begin{enumerate}
    \item A novel approach for generating adversarial input through the use of controlled morphological operations, which is then rigorously tested on a U-Net model for inland water body segmentation. Our proposed workflow is shown in Figure~\ref{fig:flowchart}.
    \item Creation of a publicly available dataset for water body segmentation, including complex geometry of river systems, with adversarial examples for robustness.
\end{enumerate}
Through our study of adversarial robustness of water body segmentation models for SAR images, this work aims to improve the reliability of the state-of-the-art deep learning models deployed in real-world geospatial applications.

\section{Related Work}
The identification of water bodies from Earth Observation (EO) images is an important step in many applications. Conventionally, a thresholding-based approach is used for water bodies identification, but the thresholds are identified experimentally based on a subset of data~\cite{4382932}. This lacks the capability to generalize, and thresholds may be different for different subsets. Tools based on spectral indices for river identification~\cite{8752013}, along with others based on a mathematical morphological-based approach, have also been proposed~\cite{klemenjak2012automatic}. However, determining the parameters of path opening and closing is challenging and can affect the results obtained. 

Recently, semantic segmentation-based approaches have been utilized to identify water bodies in EO images. Semantic segmentation aims to assign a class label to each pixel in an image. In the past decade, deep learning-based approaches have obtained excellent results in semantic segmentation for various applications. For instance, several deep learning based approaches have been utilized for semantic segmentation of water bodies~\cite{8258373} \cite{Jinsong2021TGRSWaterSeg}, including the usage of networks such as FCN-8s~\cite{long2015fully}, \texttt{tiramisu}~\cite{jegou2017one}, and \texttt{pix2pix}~\cite{isola2017image}, which is a GAN which has been adapted for the segmentation task. In these supervised learning based segmentation approaches, the training images must be manually labelled by a human. This is a time-consuming and error-prone task.  Unsupervised learning methods for segmentation on Sentinel-1 SAR imagery~\cite{OBIDA2019101910} have also been explored for river identification.  Other interesting works include a Dynamic Snake Convolution~\cite{qi2023dynamic}, which can be used to segment tubular structures such as blood vessels and roads and may be adapted for river segmentation. However, these segmentation approaches focus mainly on accurately identifying water bodies. In comparison, the proposed work studies the performance of segmentation models (primarily U-Net) in water body identification \textit{under adversarial attacks} from EO images. Our work for semantic segmentation is inspired by DeepRivWidth~\cite{verma2021deeprivwidth}, which uses two primary segmentation models, namely U-Net~\cite{ronneberger2015u} and DeepLabV3+~\cite{chen2018encoder} for segmentation of Sentinel-1 SAR images. 

There exist multiple datasets for water body identification, along with various sources in the form of shapefiles for water body mapping~\cite{sen1floods119150760, warmedinger2023new, hijmans2001computer, yamazaki2019merit, s1s210321672}. Examples include the~\textit{Sen1Floods11} dataset \cite{sen1floods119150760}, which uses raw Sentinel-1 imagery, and classifies flood water and permanent water, for flood mapping at a global scale, and the Hydrosheds Dataset~\cite{warmedinger2023new}, which provides hydrographic data products including catchment boundaries, rivers, lakes, etc. Other examples include DIVA-GIS~\cite{hijmans2001computer}, which provides comprehensive shapefiles for river basins that can be used for ground truth generation, and MERIT Hydro~\cite{yamazaki2019merit}, which provides high-resolution hydrography maps using elevation data and water body datasets such as G1WBM (Global 1-second Water Body Map), GSWO (Global Surface Water Occurrence), and OpenStreetMap.

However, we determined that these publicly available curated datasets are not suited for our task, either due to insufficiency of information in these datasets or unrecorded changes in river geometry. For the former, we found that these datasets did not contain the entire information required in our work, for instance, a few shapefiles only contained the direction that the river flows in, without the information about the river width. For the latter, we found inconsistencies in cases where the river changed its course over time. We observed that the S1-S2 Water Dataset~\cite{s1s210321672} is the best suited for the semantic segmentation task, due to the highly accurate and quality-controlled water masks, and we adopt their methodology for dataset generation in our work. This work uses the NDWI index calculated from Sentinel-2 images to create water masks. 

Many of the previously discussed works use human annotation to generate the ground truth for all the images. Not only is this a laborious and time-consuming task, but it is also highly error-prone. There have been several works aimed at studying the error types and their effects on the model performance. Most errors from a human annotator occur at the \textit{physical} boundaries of class labels in spatial applications, which may be classified as biased or unbiased errors~\cite{heller2018imperfect}. Biased errors may be repeated under similar conditions by the annotator due to human limitations in comprehension and discernibility of the classes, while unbiased errors stem from circumstances (such as a slip of a hand), and are less likely to be repeated in relabeling. 

There has been a focus on the effect of labels in semantic segmentation and general supervised learning techniques in the state of the art in deep learning, including the study of algorithms aimed at the determination of label quality for images annotated by humans~\cite{lad2023estimatinglabelqualityerrors}, along with the effect of introducing label errors in test sets to degrade the performance of the model~\cite{northcutt2021pervasive}. Studies have also shown that boundary-localized errors pose a fundamentally different and much more challenging problem than non-boundary localized errors~\cite{heller2018imperfect}. 

This motivates our novel contribution of using \textit{morphological label noise} to simulate errors produced at boundary points of the inland water bodies in the image. Our proposed strategy uses morphological operations in a controlled manner to generate adversarial inputs from this initial dataset, which is also discussed later. This particular methodology of corruption is referred to as data poisoning, and it can effectively reprogram algorithms with potentially malicious intent.

Adversarial attacks are generally classified into white- and black-box attacks, based on the level of access granted to the attacker~\cite{ZHU2024128512}. In white-box attacks, the attackers obtain all information about the target model, including its structure, parameters, etc, and they generally directly interact with the model. On the other hand, for black-box adversaries, the attacker only needs access to the output information of the target model to launch an attack. Black-box attacks are generally more realistic attacks, as the adversary may not always have access to the internal workings of the system. Data poisoning attacks can be either black- or white-box, as they involve changes to the training data itself. Also, depending on the nature of the changes made, they can be considered either a white- or a black-box change~\cite{lin2021mlattackmodelsadversarial}.

In the field of adversarial attacks and defenses, significant research has been conducted to safeguard deep neural networks (DNNs) from adversarial manipulations that can degrade their performance. One notable work is the introduction of Vector Quantization U-Net (VQUNet)~\cite{He_2024}, which addresses the vulnerability of DNNs by employing a novel noise-reduction mechanism to combat adversarial attacks. VQUNet leverages discrete latent representation learning through a multi-scale hierarchical structure to both reduce adversarial noise and reconstruct the original data with high fidelity.  Gu~\etal~\cite{10.1145/3468920.3468933} present a domain adaptation approach for semantic segmentation by leveraging the NDWI index to enhance the edge detail in pixel-level classifications. NDWI improves segmentation accuracy by addressing data distribution differences between source and target domains. Their adversarial learning framework aligns segmentation outputs, similar to adversarial attacks in NLP and vision, where data discrepancies are exploited to compromise model robustness and consistency across domains. 

There have also been several works that specifically target data poisoning attacks. One such work proposes a poisoning technique similar to ours, where the adversary targets labels in a multi-class setting and is constrained under a label-flipping budget~\cite{9897807}, which is an idea we employ in our work as well. There also exist targeted poisoning attacks, which target one specific test instance, rather than the entire training set~\cite{shafahi2018poison}. Our work targets the training data instances via a controlled corruption mechanism, which uses morphological operations, and hence, is different from previous studies.

\section{Method}
\begin{figure}[tbp]
    \centering
    \subfloat[]{\label{fig:mask1}
        \includegraphics[width=0.7\linewidth]{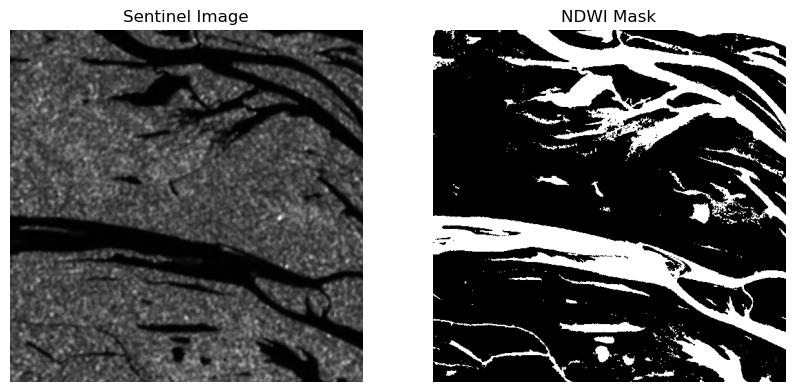}
    }\\[0.5cm] 
    \subfloat[]{\label{fig:mask2}
        \includegraphics[width=0.7\linewidth]{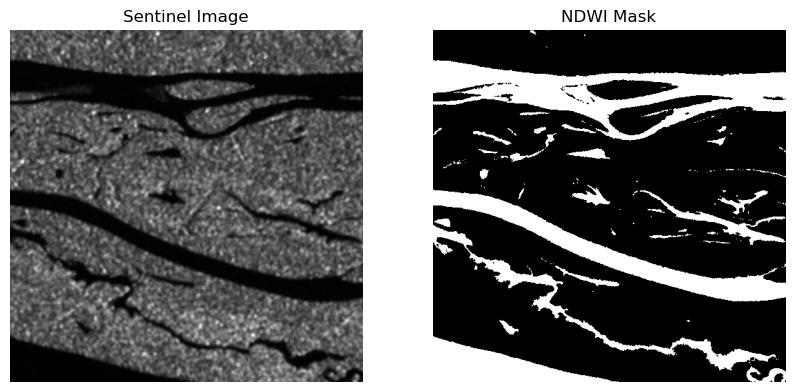}
    }
    \caption{Examples of Sentinel-1 SAR images of the Amazon River Basin (left) in our dataset, along with their generated binary water masks (right).}
    \label{fig:combined}
\end{figure}

\begin{figure}[tbp]
    \centering
    \includegraphics[width=\linewidth]{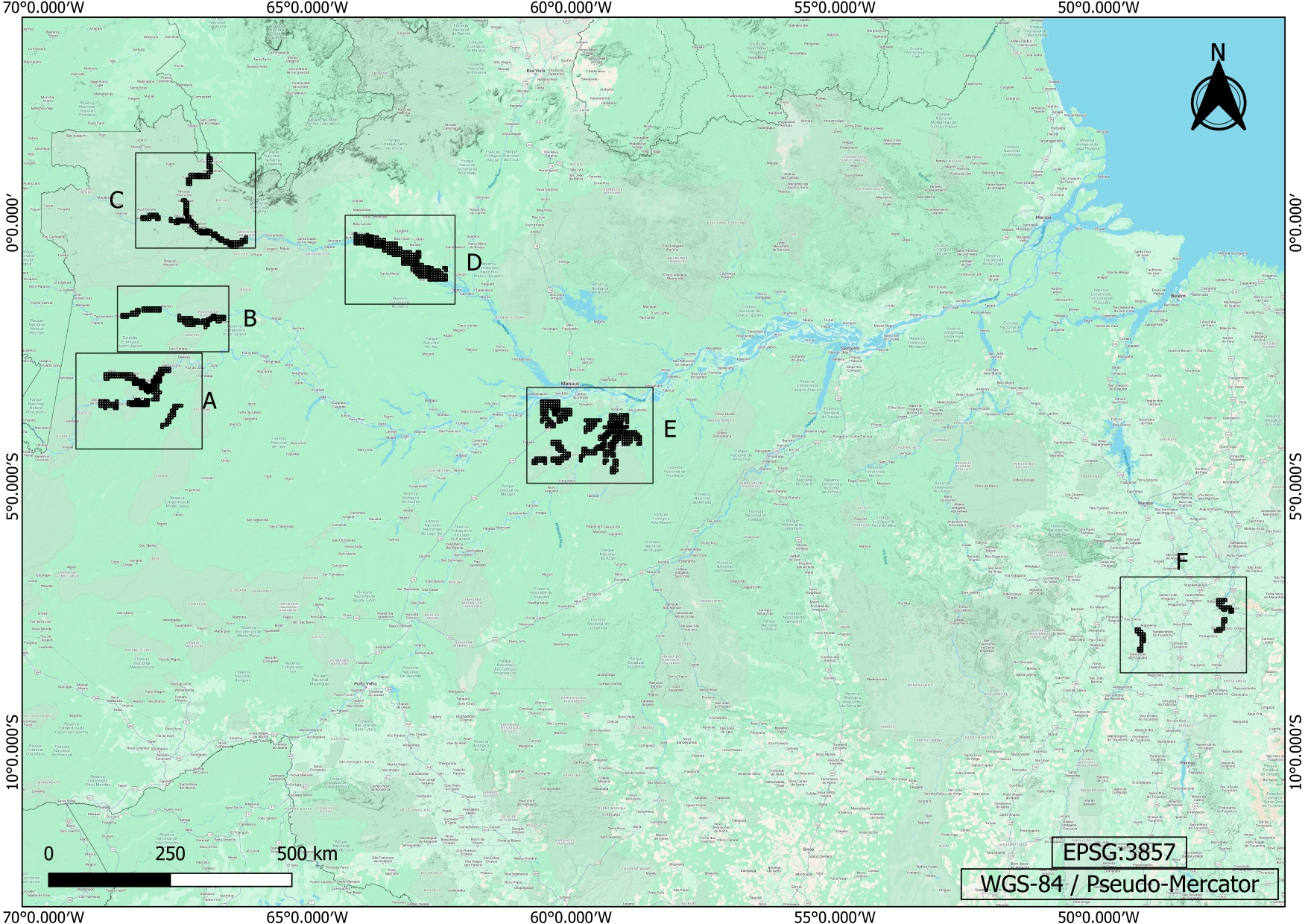}
    \caption{Regions of the Amazon River Basin used as images in our curated dataset for both training and testing. The images are taken from six non-overlapping regions, marked as regions A to F, capturing the complexity of the geometry of the river system. This image has been generated using QGIS.}
    \label{fig:regions}
\end{figure}

\begin{figure*}
    \centering
    \includegraphics[width=\linewidth]{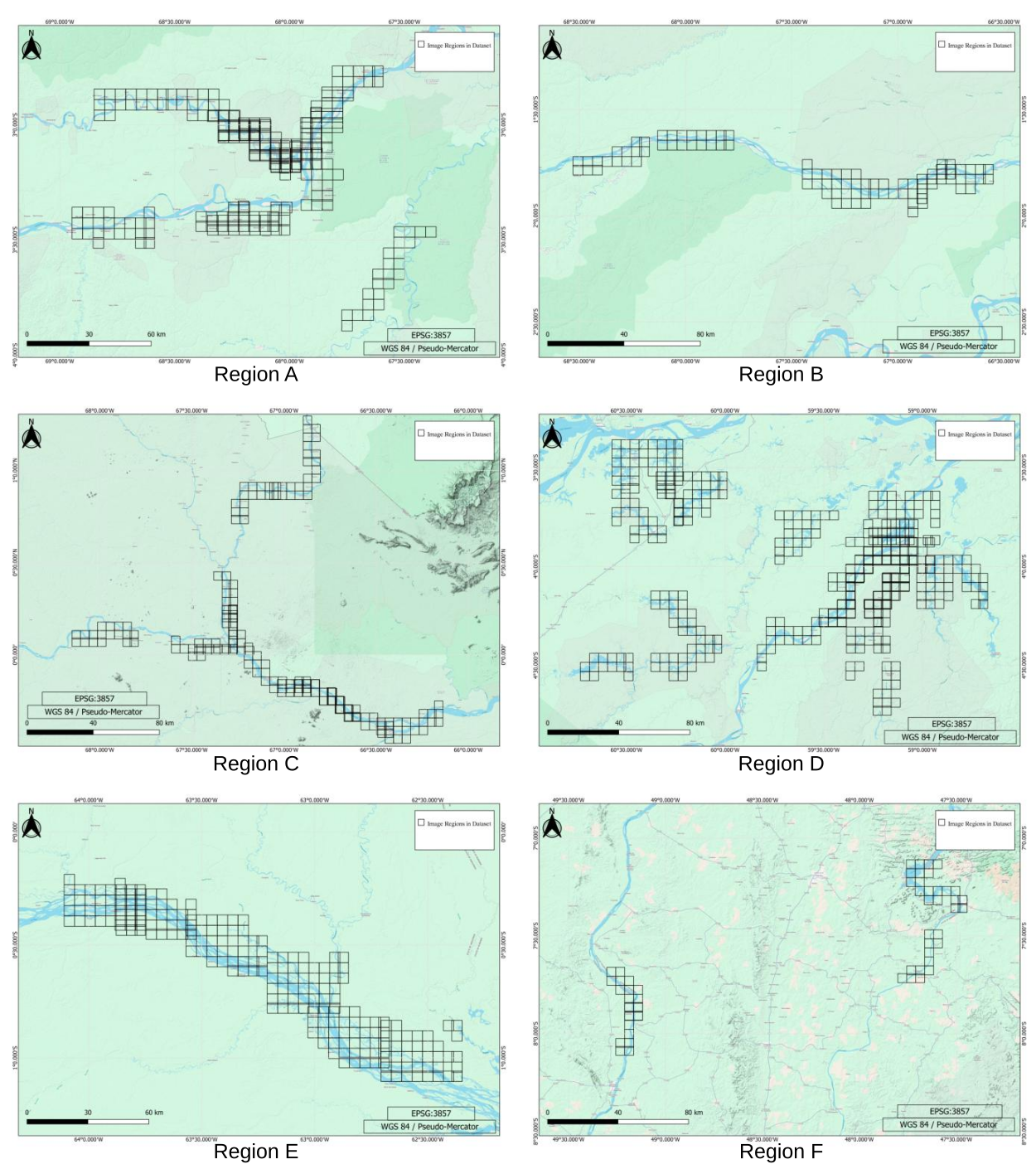}
    \caption{Zoomed-in version of regions A to F from Figure~\ref{fig:regions}, where the rectangles indicate the images used in our curated dataset. These images have been generated using QGIS.}
    \label{fig:regions_zoom}
\end{figure*}

This section describes the creation process of our proposed dataset, the segmentation models trained and tested for the dataset, and our novel simulation of adversarial attacks during training. We propose a two-stage process for the ground truth generation of water masks for our curated dataset, as shown in Figure~\ref{fig:combined}. The first stage involves the acquisition and preprocessing of Sentinel-1 images. The second stage involves the usage of the corresponding Sentinel-2 images from which water masks are obtained using NDWI indices and thresholding.

\subsection{Dataset Generation}
We selected Sentinel-1 SAR data for our case study of the Amazon River Basin from platforms like the Sentinel Scientific Data Hub and Copernicus Open Access Hub. The Amazon River Basin is an area of interest due to the large area coverage of 7,000,000 km\textsuperscript{2}, with a widespread complex river system. These river structures provide sufficient variety in the training and testing data, as shown in Figure~\ref{fig:samples}. Also, the physical structures of the river and its channels are of complex geometry, which provides several regions where human error could occur while annotating. The chosen images cover 25,194.73 km\textsuperscript{2} of the basin. The geographic locations of the images in our dataset are shown in Figures~\ref{fig:regions} and~\ref{fig:regions_zoom}.

Sentinel-1 GRD products are obtained using the Alaska Satellite Facility (ASF) API, following which we apply a sequence of several standard preprocessing steps:
\begin{enumerate}
    \item The first step is the Orbit File Application. 
    \item This is followed by Radiometric calibration~\cite{freeman1993radiometric} \cite{inproceedings}, to normalize the images under similar atmospheric and illumination conditions. 
    \item Speckle filtering \cite{lee1994speckle} is used to create a denoised image with suppressed noise. 
    \item Finally, Range Doppler Terrain correction~\cite{cumming2005digital} is performed to correct the geometric distortions based on a Digital Elevation Model (DEM). 
\end{enumerate}

To augment the dataset, we extracted images of size 512$\times$512 from the Sentinel-1 products. This entire process resulted in a total of 1,263 images.

Our initial approaches for ground truth generation involved the use of predefined shapefiles, which contain water networks. We experimented with several predefined shapefiles, but none of them generated accurate water masks suited for the segmentation task in our work. This was either due to a change in the river structure over time, or the shapefiles lacked sufficient information about width, and only contained information about the water body structure.

Finally, we implemented a widely-adopted approach using the NDWI index~\cite{s1s210321672}. The multispectral images from Sentinel-2 were obtained using Google Earth Engine. For each image, we clip out the region corresponding to the range of latitude and longitude covered by each image, and then, we calculate the NDWI index. This index is used to identify the water regions from the land regions, using a suitable threshold, for each image. This resulted in highly accurate water masks. Examples of the SAR images and their corresponding masks are shown in Figure~\ref{fig:combined}. The entire curated dataset (with correct ground truths and adversarial examples) is published at~\url{https://figshare.com/articles/dataset/Adversarial_Attack_Erosion_and_Dilation_Dataset_for_Amazon_River_Basin/28784405?file=53625587}.

\subsection{Models for Segmentation}\label{sec:method-models}

We trained a baseline model using a vanilla U-Net~\cite{ronneberger2015u} architecture, using our dataset with the accurate, \ie~uncorrupted, ground truth labels. Our goal with this initial setup was to establish a baseline model to benchmark the performance of the model on our dataset. The U-Net model has an encoder with four levels, starting with 64 filters and doubling the number at each level, reaching 512 filters. After downsampling, the bottleneck layer has two convolutional layers with 1024 filters and a dropout layer. The decoder then upsamples the feature maps back to the original resolution. The decoder also applies residual convolutional blocks to refine feature extraction. We then apply a simple threshold of 0.5 on the output of U-Net to segregate land and water pixels and generate a binary mask.

We then compare the performance of the baseline model with the state-of-the-art segmentation models on our dataset, namely, the DeepLabV3 model with a ResNet-50 backbone~\cite{chen2017rethinkingatrousconvolutionsemantic}, the SegFormer model (b-4-sized)~\cite{xie2021segformer} and the Mask2Former model with a Swin backbone~\cite{cheng2022masked,liu2021swin}. 

\subsection{Simulation of Adversarial Attacks in Training}
Erosion and dilation are two morphological operations widely used in image processing, which we use to generate controlled corruption of data in the annotated mask. Our rationale is that an insufficiently skilled human may introduce such errors in the land-water boundaries during manual annotation, especially in the presence of complex and intricate geometry of hydrological structures. In a morphological operation, the value of each pixel in the output image is based on a comparison of the corresponding pixel in the input image with its neighbors. Thus, the \textit{effect} of the morphological operation is determined by the kernel size, where the kernel function implements the operation and the kernel size determines the size of the local neighborhood.

Morphological operations are primarily suited for adding or removing pixels at the class boundaries, and hence are an ideal choice for our investigation into boundary-localized errors. Hence, we used them in the simulation of human errors for the subsequent adversarial experimentation. Figure~\ref{fig:flowchart} shows the methodology adopted for these experiments.

We also introduce randomness in the kernel sizes used for corruption, to simulate the real-world non-uniformity in the corruption. The stochastic process of selecting the kernels is essential when we need the model not to be biased in its learning. The chosen size depends on the percentage of white pixels (the pixels with mask value 1) in the image. For the images that have a white pixel percentage below 0.2, we give higher probability for choosing the kernel of size 7$\times$7 and lower probability for those with 5$\times$5 and 3$\times$3. We give the probability of 0.7 to that kernel and 0.15 to each other kernel. The exact opposite happens for dilation, we give high probability for 3$\times$3 and lower probability values for the other two kernels when the white pixel percentage is high. Finally, we utilize the following levels of corruption, given the image sizes are $512\times512$: \{$2\%$, $12\%$, $15\%$, $17\%$, $20\%$, $25\%$, $30\%$\}. 

With the model trained on correct masks, we now introduce controlled corruptions to the masks by using erosion and dilation, as these operations present examples of potential human errors in the precise delineation of land-water boundaries. Thus, these controlled corruptions are \textit{morphological label noise}, and this process of corruption is effectively a \textit{data poisoning} procedure.

Our proposed methodology for corruption is as follows --
\begin{enumerate}
\item Out of the train and the validation images, we select one-third of the images for corruption using erosion, one-third using dilation, and one-third is left uncorrupted.
\item For an experiment, we then set a threshold on the number of corrupted pixels allowed.
\item Erosion and dilation are applied to the corresponding masks iteratively, till either a pre-defined $N$ iterations are over, or the number of corrupted pixels crosses the threshold, following which we choose the previous highest corrupted version of the mask. In our work, we selected $N$=100.
\end{enumerate}

\subsection{Implementation}
The data points are divided into 883 training, 127 validation, and 253 test samples. The U-Net model is trained for 20 epochs, for this experiment and across all subsequent experiments. We used a pre-trained implementation of DeepLabV3 with ResNet-50 backbone, and finetuned it over our dataset, for 50 epochs (with an Early Stopping condition to prevent overfitting). We used SegFormer (b-4-sized) pretrained on Imagenet~\cite{5206848}, and finetuned for 50 epochs on our dataset. Finally, for Mask2Former, we used the large-sized version with Swin Transformer, pretrained on the Cityscapes dataset~\cite{cordts2016cityscapes}, and finetuned it for a total of 50 epochs on the dataset. 
We run all the models on a system with an AMD Ryzen 9 5950X 16-core processor, connected to an NVIDIA RTX A4000 graphics card. Our code is available at~\url{https://github.com/GVCL/IWSeg-SAR-Poison.git}.

\section{Experiments and Results}
\begin{table}[tp]
    \centering
    \caption{Performance of the various models on uncorrupted/clean data.}
    \label{tab:models}
    \begin{tabular}{|c|c c c c c c c|}
    \hline
    & \bf Dice && \bf Prec-
    && \bf F1 & \bf Specif- & \bf Accu- \\
    \bf Model & \bf Coeff. & \bf IoU & \bf ision
    & \bf Recall & \bf Score & \bf icity & \bf racy \\\hline

    \bf U-Net~\cite{ronneberger2015u} & 0.770 & 0.823 & 0.896
    & 0.869 & 0.865 & 0.971 & \textbf{0.955} \\
    \bf (Baseline) &&& &&&& \\\hline

    DeepLabV3 & 0.770 & 0.819 & 0.891
    & 0.865 & 0.864 & 0.965 & \textbf{0.953} \\
    \cite{chen2017rethinkingatrousconvolutionsemantic} &&& &&&& \\\hline

    SegFormer & 0.770 & 0.829 & 0.908
    & 0.868 & 0.870 & 0.975 & \textbf{0.957} \\
    \cite{xie2021segformer} &&& &&&& \\\hline

    Mask2- & 0.774 & 0.827 & 0.894
    & 0.872 & 0.869 & 0.973 & \textbf{0.957} \\
    Former~\cite{cheng2022masked} &&& &&&& \\\hline
        
    \end{tabular}
\end{table}

\begin{figure}[htbp]
    \centering
    \subfloat[Output of the baseline U-Net model~\cite{ronneberger2015u}\label{fig:unet10}]{
        \includegraphics[width=0.8\linewidth]{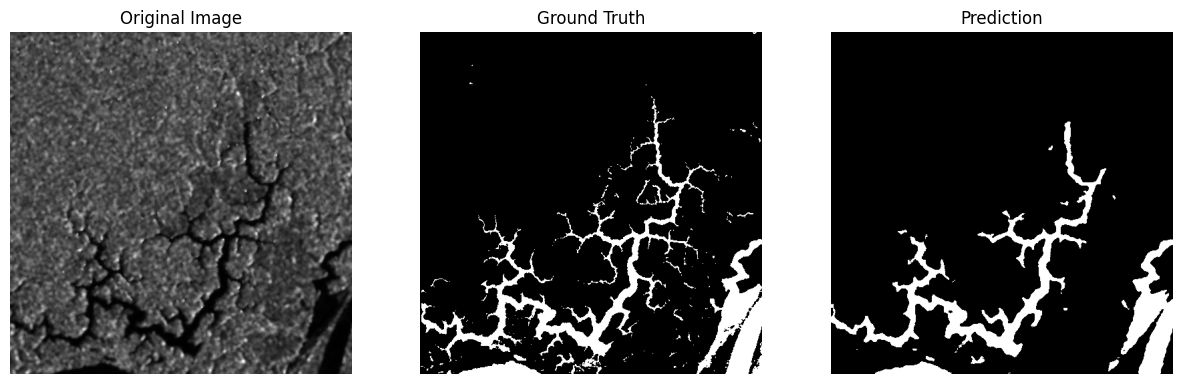}
    }\\
    \subfloat[Output of the DeepLabV3 model~\cite{chen2017rethinkingatrousconvolutionsemantic} \label{fig:deeplab10}]{
        \includegraphics[width=0.8\linewidth]{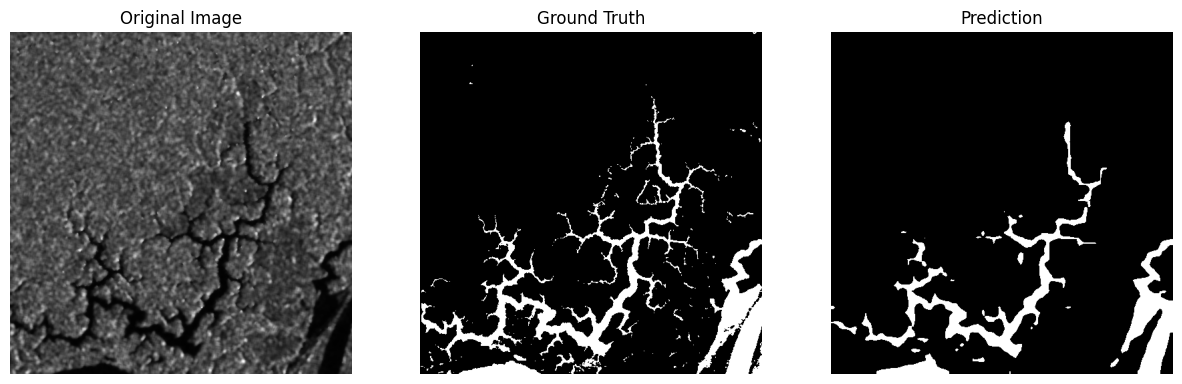}
    }\\
    \subfloat[Output of the SegFormer model~\cite{xie2021segformer}\label{fig:segformer10}]{
        \includegraphics[width=0.8\linewidth]{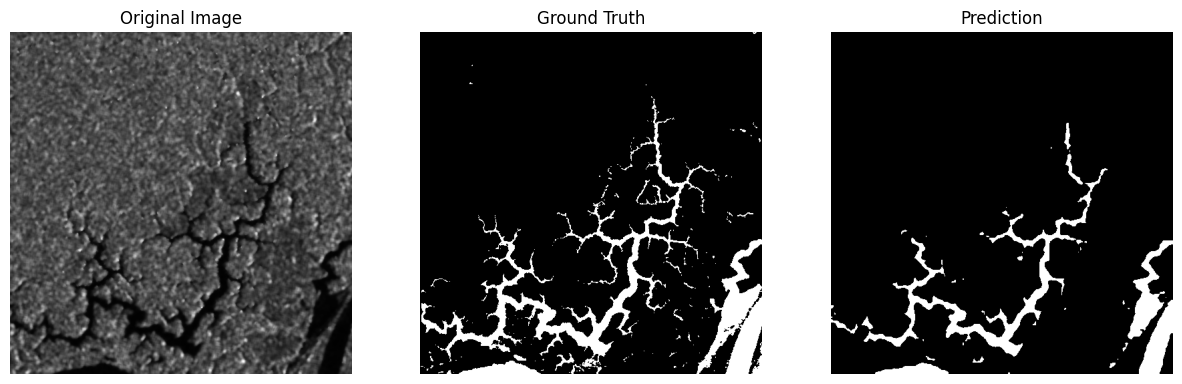}
    }\\
    \subfloat[Output of the Mask2Former model~\cite{cheng2022masked} \label{fig:mask2former10}]{
        \includegraphics[width=0.8\linewidth]{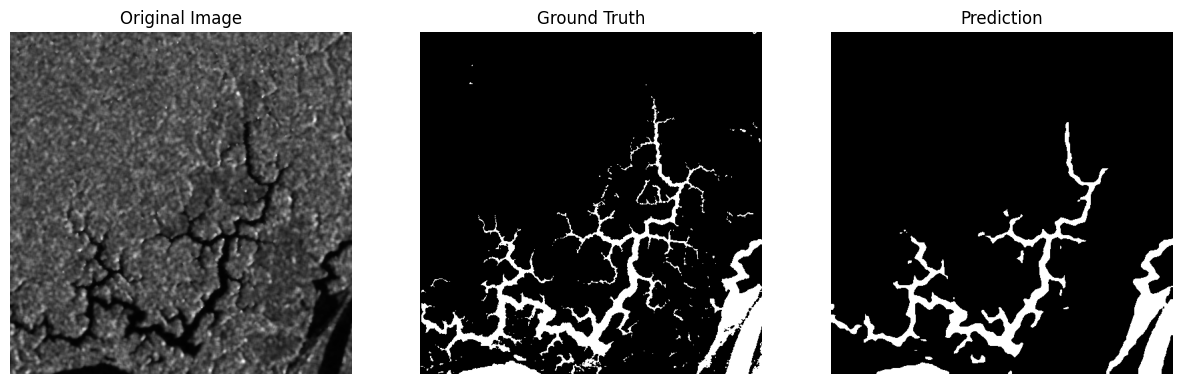}
    }
    \caption{Outputs of the baseline and state-of-the-art models using uncorrupted masks on a data point where all the models perform poorly.}
    \label{fig:outputs_clean}
\end{figure}

\begin{figure}[tbp]
    \centering
    \includegraphics[width=0.7\linewidth]{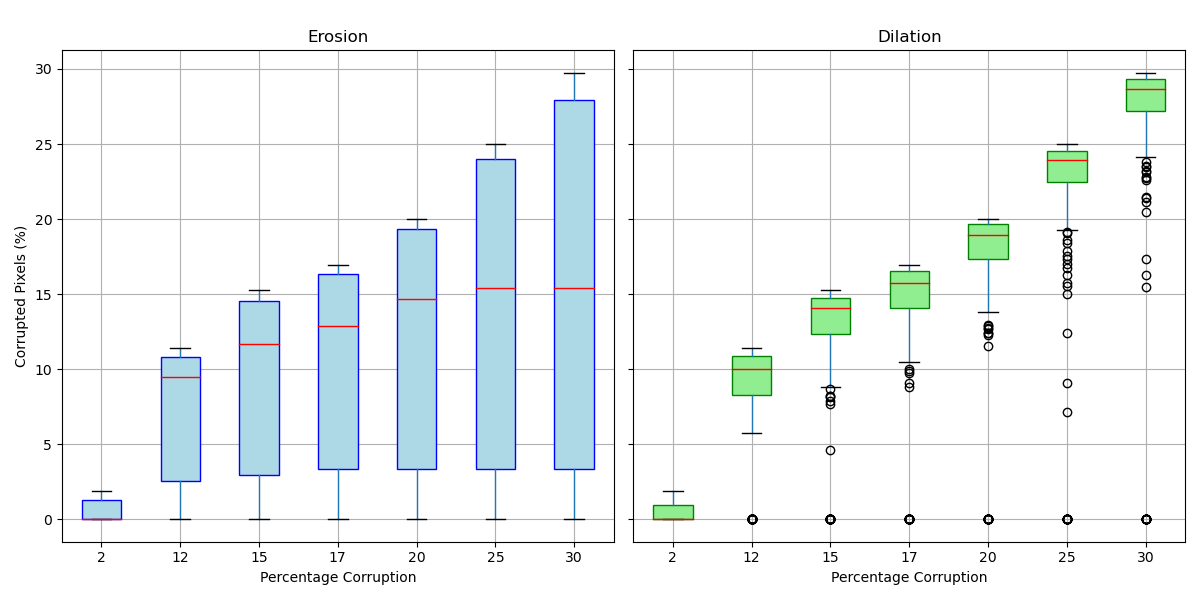}
    \caption{A box plot showing the range of actual corruption in the training data for different corruption levels.}
    \label{fig:corruption_boxplot}
    \vspace{1em}
    
    \includegraphics[width=0.7\linewidth]{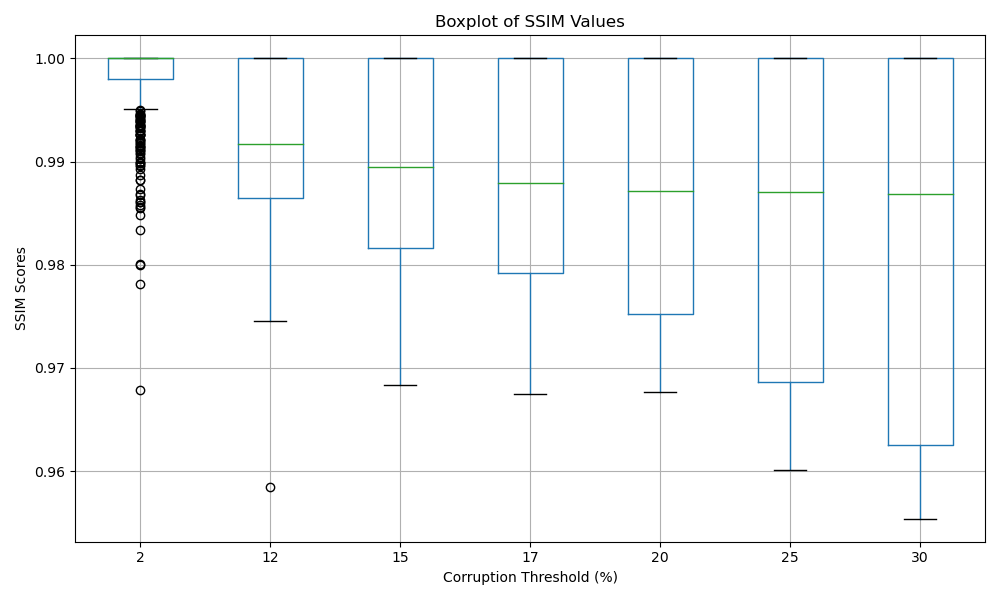}
    \caption{A box plot showing the SSIM values across all images between the uncorrupted mask and the corresponding corrupted masks for various levels of corruption.}
    \label{fig:ssim_boxplot}
\end{figure}

In this section, we present the results of all of the models run on our uncorrupted dataset, and also those of our baseline U-Net model on the dataset corrupted by the adversarial attacks.

\subsection{Segmentation Results with the Uncorrupted Dataset}

We present the results of state-of-the-art segmentation models (U-Net, DeepLabV3, SegFormer, Mask2Former) on our dataset (uncorrupted), which helps us assess the quality of the dataset. Table~\ref{tab:models} shows the performance of the models on the dataset. The performance of each of the models saturates at a per-pixel accuracy of $\sim$0.955. Moreover, the IoU for all these models is comparable.

We further investigate the samples that none of the models are able to segment properly. A few such \textit{failure cases} are discussed in Section~\ref{Discussion}, where either due to incorrect thresholding, or due to unavailability of a complete Sentinel-2 image, the ground truth itself is incorrect. The incorrectness of the ground truth consequently leads to a higher level of segmentation inaccuracy.

However, one consistent failure observed across the model is its inability to fully delineate narrow tributaries in the image. An example of one such failure case is shown in Figure~\ref{fig:outputs_clean}. Overall, we observe that all our selected models face issues in delineating the very narrow tributaries and are only somewhat correct, with varying levels of success.

\subsection{Adversarial Experimentation}

\begin{figure}[tbp]
    \centering
    \includegraphics[width=0.5\linewidth]{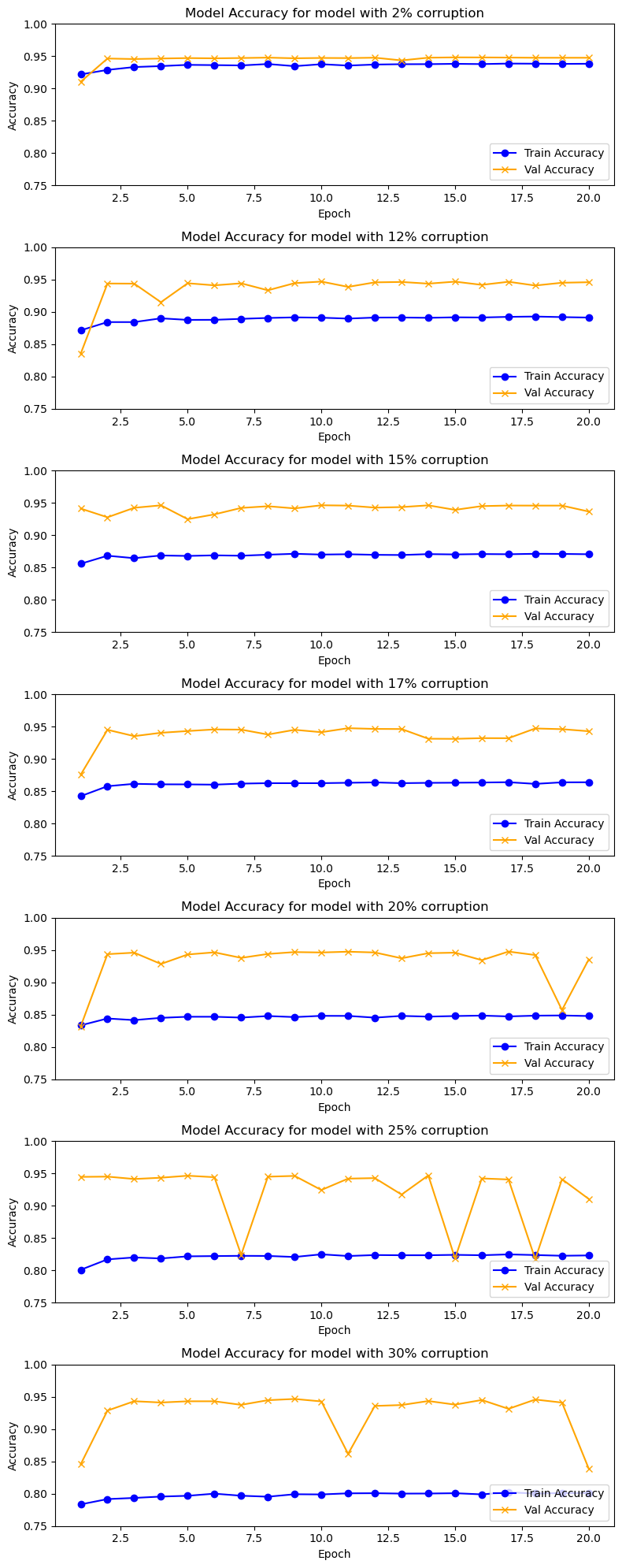}
    \caption{Training and validation accuracy plots for different corruption levels while using controlled morphological noise for corruption.}
    \label{fig:epoch_plots}
\end{figure}

It is to be noted that although we set a limit on the number of corrupted pixels, in practice, we may not be able to reach that limit, especially in the erosion operation. Erosion reduces the number of white pixels, and hence, whenever it reaches a state of insufficient white pixels, it does not allow further increase in the number of corrupted pixels.

This saturation of corruption is demonstrated in Figures~\ref{fig:corruption_boxplot} and~\ref{fig:ssim_boxplot}. Figure~\ref{fig:corruption_boxplot} shows the box plot of the actual number of corrupted pixels across all training images for different corruption levels, where the corruption is by erosion, and a similar box plot for dilation. We use the Structural Similarity Index (SSIM) score to quantify the similarities between two images. Figure~\ref{fig:ssim_boxplot} shows the distribution of the SSIM scores between the uncorrupted ground truth and the corresponding corrupted ground truth, computed across all images and for different corruption levels. We observe that the number of pixels corrupted for images undergoing erosion becomes stagnant as we increase the corruption levels, with the median and the lower quartile barely changing beyond $20\%$ corruption. The dilation plots, on the other hand, show the expected trends. The same is true for the SSIM values between the uncorrupted masks and their corresponding corrupted masks. Thus, we conclude that, in our approach, the corruptions may be limited by design, as the final count of corrupt pixels for an image may be less than that expected in the targeted level of corruption.

\subsection{Results on Corrupted Dataset and Discussion} \label{Discussion}

\begin{table}[tp]
    \centering
    \caption{Performance of the baseline U-Net model on our dataset at various corruption levels}
    \label{tab:unet_res}
    \begin{tabular}{|c|c c c c c c c|}
    \hline
    \bf Corruption & \bf Dice && \bf Prec-
    && \bf F1 & \bf Specif- & \bf Accu- \\
    \bf Level & \bf Coeff. & \bf IoU & \bf ision
    & \bf Recall & \bf Score & \bf icity & \bf racy \\\hline

    0\% & 0.770 & 0.823 & 0.896
    & 0.869 & 0.865 & 0.971 & \bf 0.955 \\
    (Clean) &&& &&&& \\\hline

    2\% & 0.775 & 0.825 & 0.896
    & 0.873 & 0.868 & 0.968 & 0.955 \\\hline

    12\% & 0.769 & 0.822 & 0.899
    & 0.864 & 0.865 & 0.975 & 0.955 \\\hline

    15\% & 0.768 & 0.819 & 0.890
    & 0.866 & 0.863 & 0.971 & 0.954 \\\hline
    
    17\% & 0.758 & 0.814 & 0.903
    & 0.854 & 0.859 & 0.979 & 0.952 \\\hline

    20\% & 0.725 & 0.790 & 0.902
    & 0.830 & 0.841 & 0.981 & \textbf{0.946} \\\hline

    25\% & 0.597 & 0.708 & 0.875
    & 0.759 & 0.768 & 0.984 & \textbf{0.920} \\\hline

    30\% & 0.253 & 0.496 & 0.818
    & 0.588 & 0.567 & 0.996 & \textbf{0.828} \\\hline
    \end{tabular}
\end{table}

Table~\ref{tab:unet_res} shows the evaluation metrics of training the U-Net model across the various levels of corruption. We get a good 95\% accuracy in segmentation. Furthermore, the performance is mostly unaffected even after increasing the corruption levels to as high as 20\%. This is surprising, as the model is able to tolerate even such high levels of corruption. We observe that the model performance does not deteriorate significantly until 25\% corruption. Hence, we investigate more into the training of the model, the qualitative predictions of the model on a particular image, and the actual corruption that the model is facing.

Figure~\ref{fig:epoch_plots} shows the per-epoch accuracy on the training and validation data for each of the corruption levels. The plots clearly indicate that until 17\% corruption, the validation accuracy does not reduce significantly. But, beyond a 17\% corruption level, the model begins to show instability in the validation accuracies. Further, we observe that the model is able to avoid overfitting, as with the increase in the corruption level, the training accuracy falls, but never the validation accuracy. 

\begin{figure*}[htbp]
    \centering
    \includegraphics[width=0.7\linewidth]{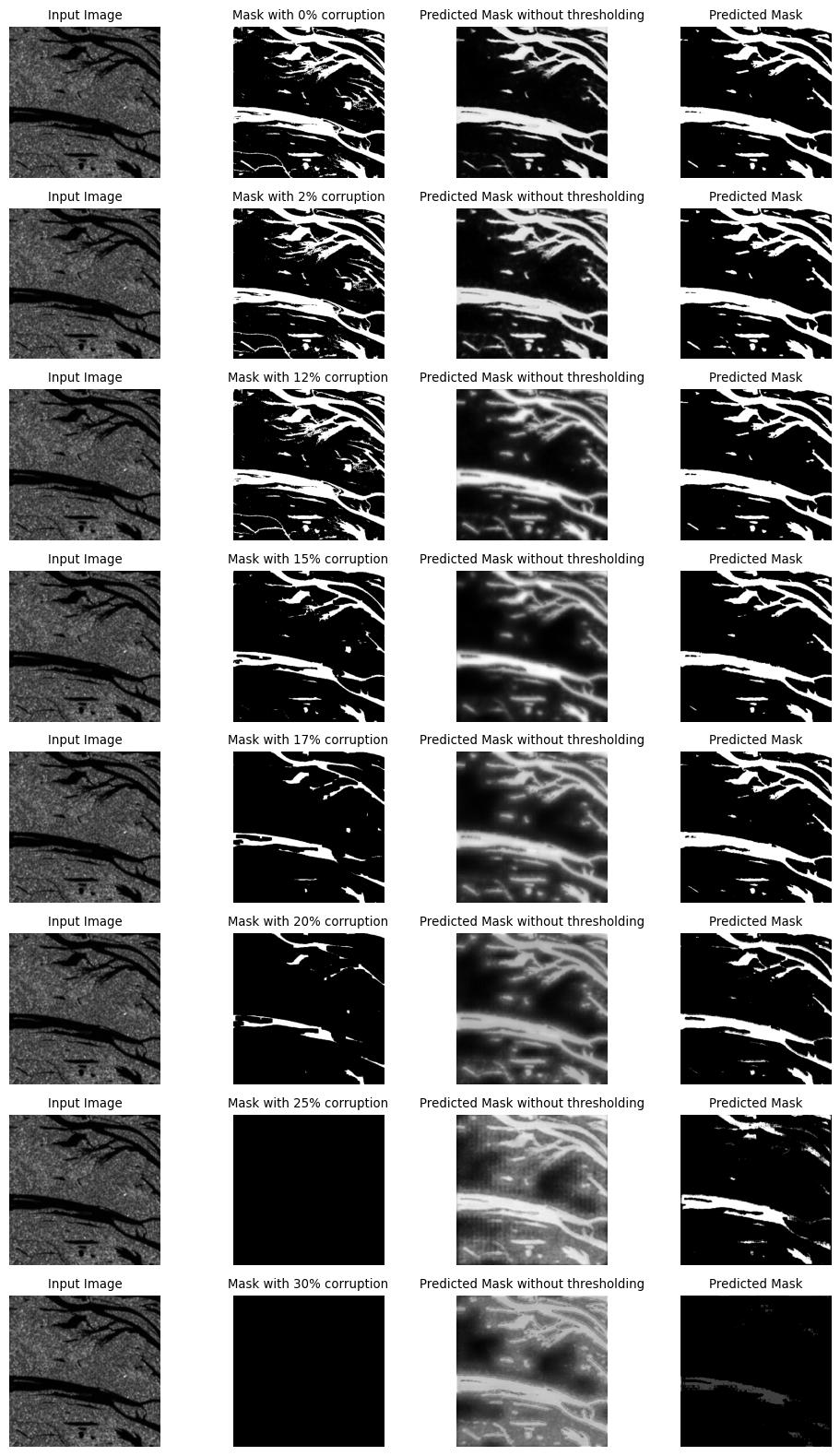}
    \caption{Qualitative results showing the effect of corruption level on the ground truth (second column from the left) and the final predicted (rightmost column) mask, for the input image shown in Figure~\ref{fig:mask1} (leftmost column). The predicted masks before applying the threshold to the U-Net model outcomes (second column from the right) get blurrier with an increase in corruption, but are robust in capturing river structure in the presence of morphological label noise.}
    \label{fig:dense}
\end{figure*}

\begin{figure*}[htbp]
    \centering
    \includegraphics[width=0.7\linewidth]{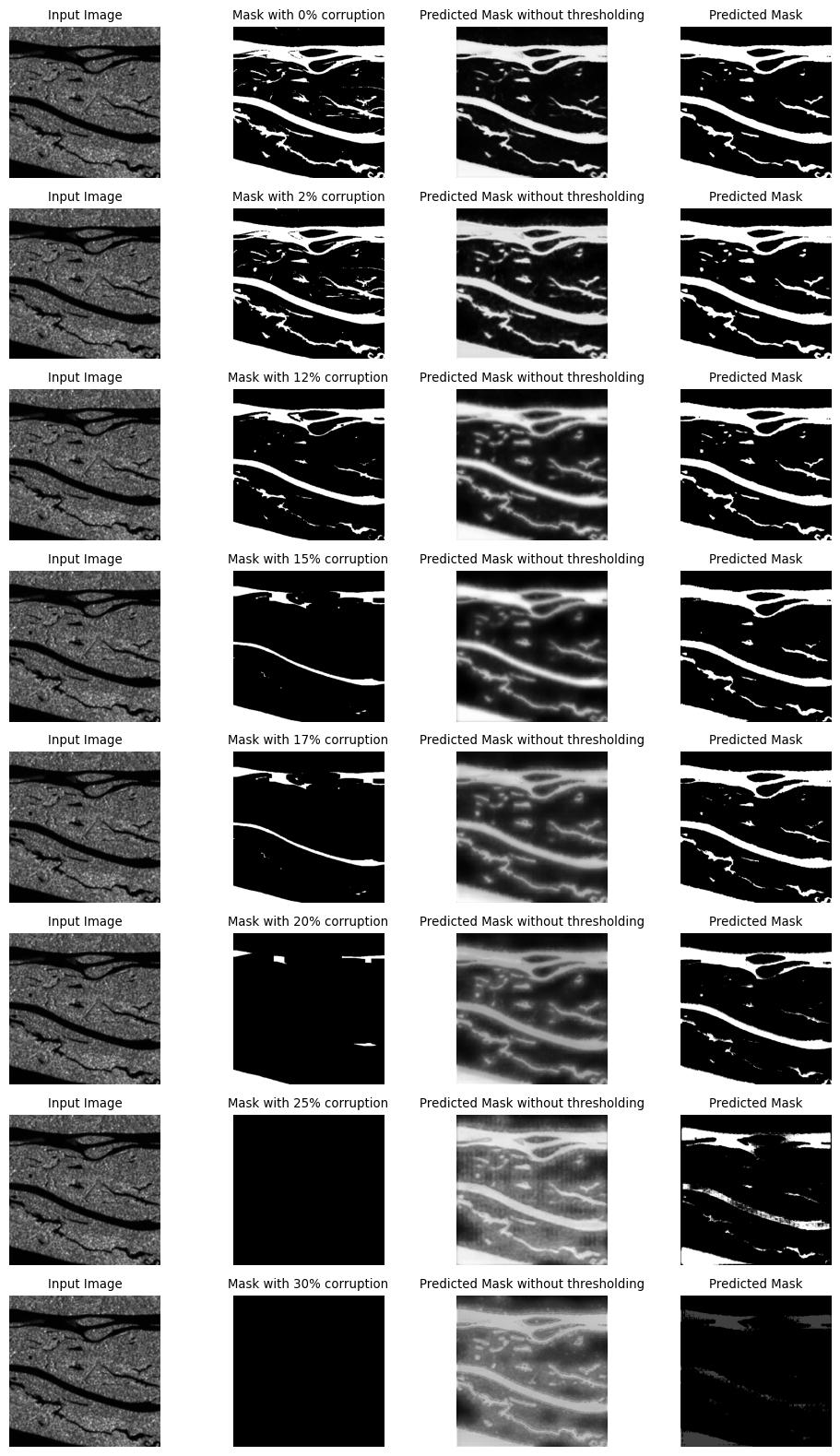}
    \caption{Qualitative results showing the effect of corruption level on the ground truth (second column from the left) and the final predicted (rightmost column) mask, for the input image shown in Figure~\ref{fig:mask2} (leftmost column). The predicted masks before applying the threshold (second column from the right) to the U-Net model outcomes get blurrier with an increase in corruption, but are robust in capturing river structure in the presence of morphological label noise.}
    \label{fig:dense1}
\end{figure*}

\begin{figure*}[htbp]
    \centering
    \includegraphics[width=0.7\linewidth]{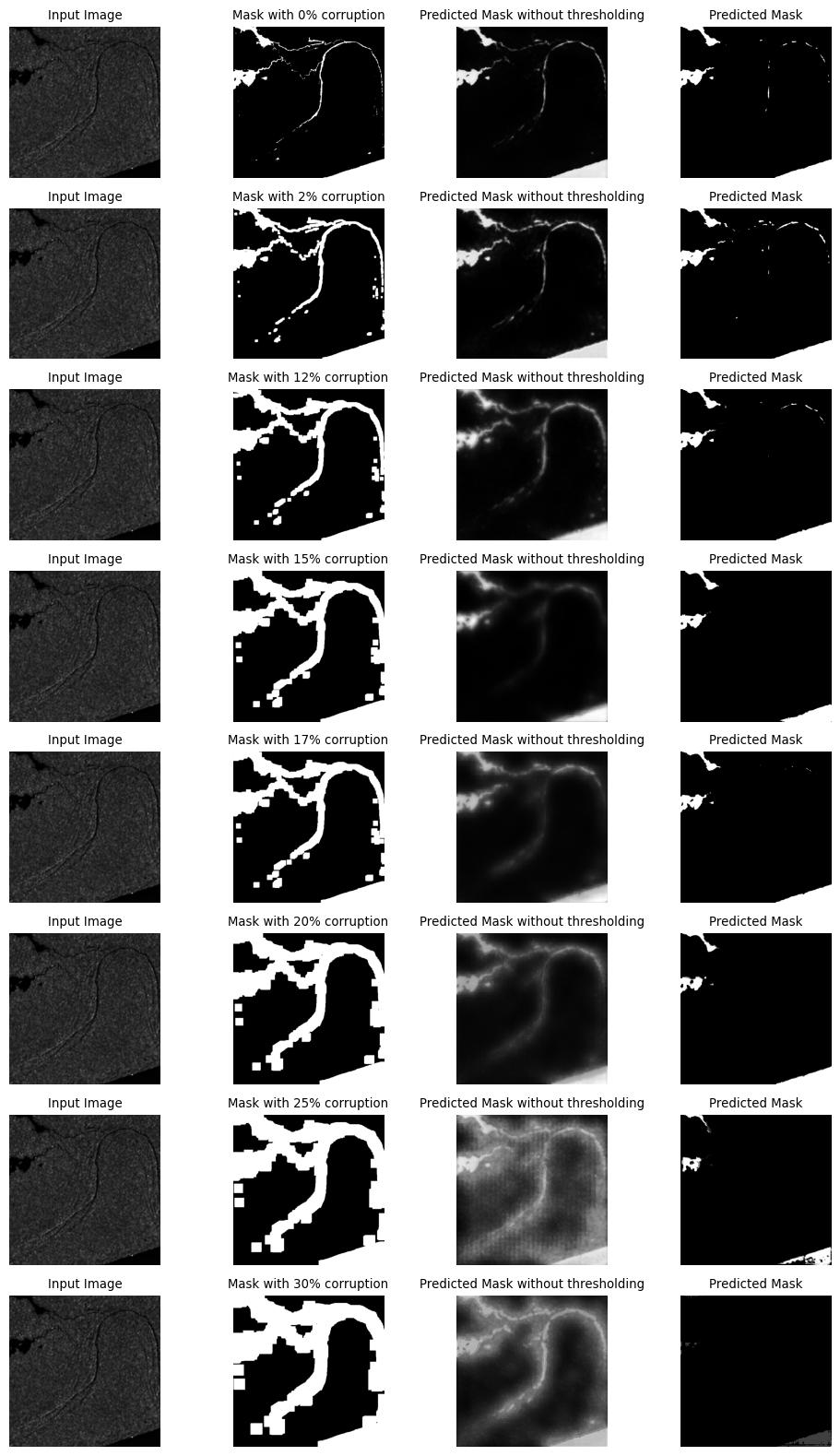}
    \caption{Results showing the effect of corruption level on the ground truth (second column from the left) and the final predicted (rightmost column) mask, for the input (leftmost column). The predicted masks before applying the threshold (second column from the right) to the U-Net model outcomes capture the river geometry, but the final outputs indicate the model's failure.}
    \label{fig:dense2}
\end{figure*}

These results are further supported by Figures~\ref{fig:dense} and~\ref{fig:dense1}. These figures show the effect of corruption on training data in two examples. They show the input image and the corresponding actual masks, which are fed to the model as inputs. These are points on which erosion has been applied, and we observe that with the increase in the corruption levels, the percentage of white pixels reduces significantly, with no remaining white pixels when the corruption is increased to 25\%. This is also an example of a data point where just increasing the corruption level does not necessarily lead to an increase in the actual corrupted pixels.

With the increase in corruption in the training data, we observe that the baseline U-Net model output, which is obtained before applying the threshold of 0.5 (explained in Section~\ref{sec:method-models}), becomes increasingly blurred. This indicates that the increase in corruption reduces the model's ability to discern land from water. However, the model remains capable of finding the river structure or geometry, and when we apply the threshold on the U-Net output, the masks predicted are still reasonably accurate. This indicates that U-Net is robust to high levels of corruption. These observations are shown in Figures~\ref{fig:dense} and \ref{fig:dense1}, where the model predictions at 20\% and 25\% corruption cases are correct in terms of the river structure, but the overall accuracy has reduced in the U-Net model outputs, obtained before the step of applying the threshold of 0.5. 

All these results show that the model is able to easily tolerate up to $17\%$ corruption, beyond which the model becomes unstable in training. The model avoids overfitting and, despite the high levels of corruption, is able to accurately obtain the river structure. Till 20\% corruption, the model is even able to accurately find the mask after thresholding, beyond which the predictions become increasingly blurred and poor.

We also include the results from a data point on which the clean U-Net model itself does not perform very accurately. One such failure case is shown in Figure~\ref{fig:dense2}. The model faces issues in accurately segmenting narrow tributaries in the image, and the model's performance further deteriorates with the increase in corruption. This example also appears to show that corruption with dilation is much more severe than erosion. This observation is also supported by Figure~\ref{fig:corruption_boxplot}, which shows that for dilation, the number of corrupted pixels continues to increase with the increase in corruption level, unlike the case with erosion. The prediction is poor even at 10\% corruption.

\begin{figure}[tbp]
    \centering
    \includegraphics[width=\linewidth]{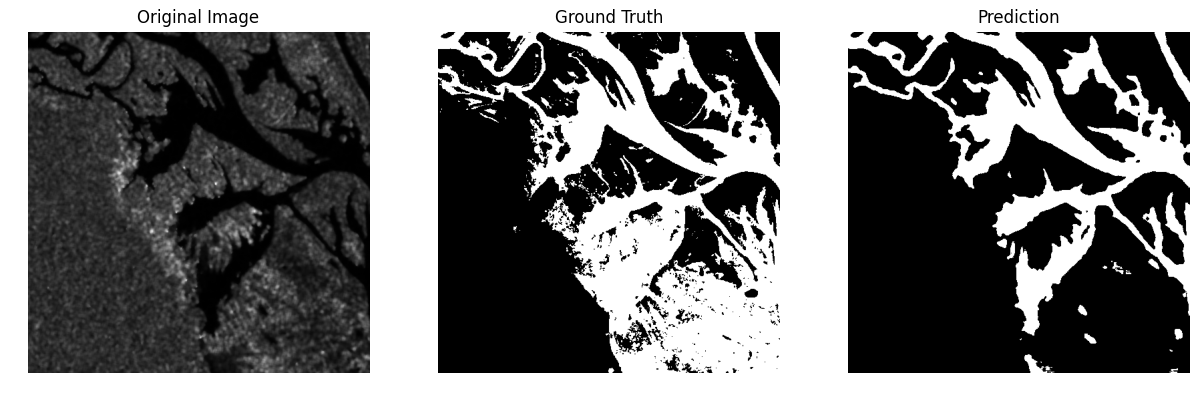}
    \caption{A data point where the ground truth is not fully accurate due to a non-optimal threshold used in ground truth generation.}
    \label{fig:corrupt_gt1}
    \vspace{1em}
    \centering
    \includegraphics[width=\linewidth]{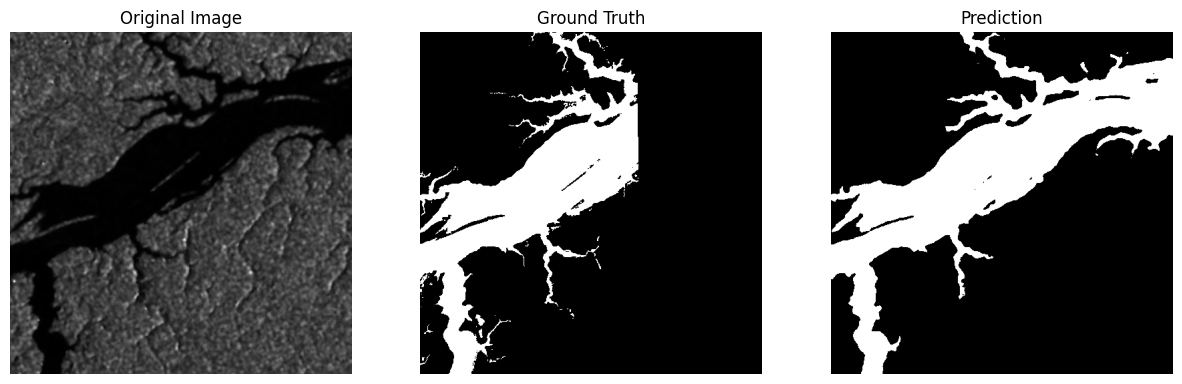}
    \caption{A data point where the ground truth is incorrect due to incomplete ground truth owing to the unavailability of the Sentinel-2 image in the specific region.}
    \label{fig:corrupt_gt2}
\end{figure}

\subsection{Observed Anomalies in the Generated Ground Truth}
We observed two different types of anomalies in the generated ground truth. Here, we report the findings of the model performance in the presence of the anomalies, which may be considered as \textit{inadvertent label noise}. These anomalies are different from the controlled morphological label noise. Thus, the previous references of ``uncorrupted masks'' in this paper meant that the masks did not have the controlled label noise, but still contained the inadvertent label noise.

The first anomaly is from the choice of the threshold for the NDWI index when generating the mask for the images. This threshold has been manually selected. At times, this threshold need not be the most optimal one for the image, as shown in Figure~\ref{fig:corrupt_gt1}. As observed, the ground truth is not fully accurate and contains regions that are not water bodies but are misclassified as water. However, we observe that the training is robust even in this case, as the mask predicted by the baseline U-Net model in the absence of label noise is qualitatively accurate when compared to the actual image. The same is also observed for all the other models. However, this anomaly can be avoided by adaptively selecting the optimal threshold for the NDWI index~\cite{CHE2025132771}, while creating the ground truth, \ie~the uncorrupted masks.

\begin{figure}[tbp]
    \centering
    \includegraphics[width=\linewidth]{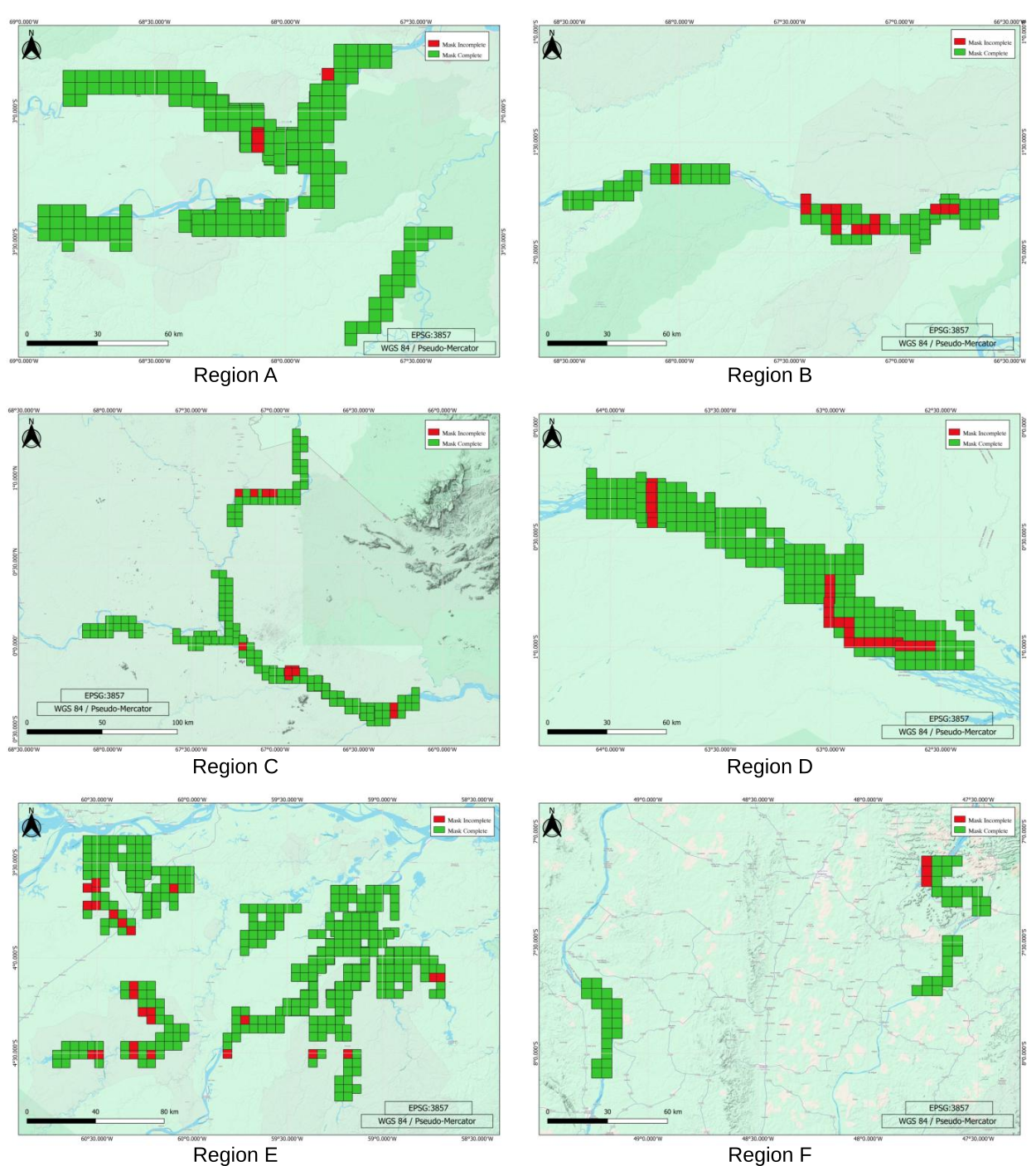}
    \caption{Images with accurate masks (green) and anomalous masks (red), where the latter is due to incomplete data in Sentinel-2 images for ground truth generation. Observing these masks superimposed on the Amazon River Basin helped identify the cause for this anomaly.}
    \label{fig:complete_masks}
\end{figure}

The second anomaly arises from incomplete data in the Sentinel-2 images. We observed that the Sentinel-2 images of certain regions were not fully available, due to which the generated mask is incomplete. An example of this anomaly is shown in Figure~\ref{fig:corrupt_gt2}. We further located such anomalies in the entire curated dataset to check the level of inadvertent corruption due to the unavailable data. We found that this anomaly is concentrated to a few regions, as shown in Figure~\ref{fig:complete_masks}. The red squares indicate the regions with anomalous images, while the green ones are accurate. We observe that almost all the inaccurate masks occur in clusters, except in Region E, thus helping us identify the cause of the anomalies. The models are robust even in this scenario, being able to accurately predict the mask when compared with the actual image, as shown in Figure~\ref{fig:corrupt_gt2}. We can see the qualitatively accurate result in this case, even though the actual accuracy value would be computed less with respect to the incorrect ground truth. 

Overall, these results show that all the models trained with uncorrupted masks are fairly robust to these anomalies in the dataset generation, highlighting their suitability for the task of inland water body segmentation. However, the only limitation of these models, as alluded to earlier, is in the accurate delineation of small tributaries. 

We chose the U-Net model for adversarial experimentation primarily due to its ability to achieve favorable performance even with relatively smaller training sets, and its ability to run faster than other models. Also, U-Net is known to be more stable to errors than other architectures~\cite{heller2018imperfect}. Future scope of work can include performing similar experiments on other segmentation architectures and comparing stability and performance. One line of work can be to observe the effects of only erosion and only dilation on the datasets, to quantify and compare the extent of degradation in model performance by the two methods. Another interesting direction of work would be to include other, more traditional modes of corruption, including rotation and translation, and observing their effects. 

\section{Conclusions}
In recent years, several semantic segmentation models have been proposed to identify inland water bodies from Synthetic Aperture Radar (SAR) images. Most of these supervised segmentation models depend on manually annotated labels. The performance of these supervised models depends on the accuracy of labels. In this study, we simulated manual annotation errors through adversarial attacks on the U-Net model to examine the model's robustness against human annotation errors. We also introduced a new method for simulating human-induced errors using morphological operations and employed a controlled corruption strategy to evaluate the performance of the U-Net model. Our testing and observations indicate that the model is robust against morphological forms of corruption and can accurately detect water structures, even in the presence of high levels of corruption, up to a certain threshold. This finding underscores the vital importance of the quality of manual annotations in influencing the performance of the segmentation model.

Additionally, we developed a new dataset for the semantic segmentation task, which can be used to benchmark the performance of segmentation models specifically for water body segmentation. This dataset includes adversarial examples that can test the robustness of the models under various levels of corruption. 

\bibliographystyle{unsrt}
\bibliography{papers_riverseg}

\begin{thebibliography}{10}

\bibitem{8258373}
Laura Lopez-Fuentes, Claudio Rossi, and Harald Skinnemoen.
\newblock {River Segmentation for Flood Monitoring}.
\newblock In {\em 2017 IEEE International Conference on Big Data (Big Data)},
  pages 3746--3749, 2017.

\bibitem{WANG2022289}
Shidong Wang, Maria~V. Peppa, Wen Xiao, Sudan~B. Maharjan, Sharad~P. Joshi, and
  Jon~P. Mills.
\newblock {A Second-order Attention Network for Glacial Lake Segmentation from
  Remotely Sensed Imagery}.
\newblock {\em ISPRS Journal of Photogrammetry and Remote Sensing},
  189:289--301, 2022.

\bibitem{OSULLIVAN2024101276}
Conor O’Sullivan, Ambrish Kashyap, Seamus Coveney, Xavier Monteys, and
  Soumyabrata Dev.
\newblock {Enhancing Coastal Water Body Segmentation with Landsat Irish Coastal
  Segmentation (LICS) Dataset}.
\newblock {\em Remote Sensing Applications: Society and Environment},
  36:101276, 2024.

\bibitem{pech2023sentinel}
Fernando Pech-May, Ra{\'u}l Aquino-Santos, and Jorge Delgadillo-Partida.
\newblock {Sentinel-1 SAR Images and Deep Learning for Water Body Mapping}.
\newblock {\em Remote Sensing}, 15(12):3009, 2023.

\bibitem{ronneberger2015u}
Olaf Ronneberger, Philipp Fischer, and Thomas Brox.
\newblock {U-Net: Convolutional Networks for Biomedical Image Segmentation}.
\newblock In {\em Proceedings of 18th International Conference on Medical Image
  Computing and Computer-Assisted Intervention (MICCAI 2015), part III 18},
  pages 234--241. Springer, 2015.

\bibitem{erfani2022atlantis}
Seyed Mohammad~Hassan Erfani, Zhenyao Wu, Xinyi Wu, Song Wang, and Erfan
  Goharian.
\newblock {ATLANTIS: A Benchmark for Semantic Segmentation of Waterbody
  Images}.
\newblock {\em Environmental Modelling \& Software}, 149:105333, 2022.

\bibitem{xia2025openearthmapsarbenchmarksyntheticaperture}
Junshi Xia, Hongruixuan Chen, Clifford Broni-Bediako, Yimin Wei, Jian Song, and
  Naoto Yokoya.
\newblock {OpenEarthMap-SAR: A Benchmark Synthetic Aperture Radar Dataset for
  Global High-Resolution Land Cover Mapping}, 2025.

\bibitem{Rahaman18082022}
Mustafizur Rahaman, Md.~Monsur Hillas, Jannatul Tuba, Jannatul~Ferdous Ruma,
  Nahian Ahmed, and Rashedur M.~Rahman and.
\newblock {Effects of Label Noise on Performance of Remote Sensing and Deep
  Learning-based Water Body Segmentation Models}.
\newblock {\em Cybernetics and Systems}, 53(6):581--606, 2022.

\bibitem{isprs-archives-XLVIII-M-3-2023-81-2023}
A.~Edpuganti, P.~Akshaya, J.~Gouthami, V.~V. Sajith~Variyar, V.~Sowmya, and
  R.~Sivanpillai.
\newblock {Effect of Data Quality on Water Body Segmentation with DeepLabV3+
  Algorithm}.
\newblock {\em The International Archives of the Photogrammetry, Remote Sensing
  and Spatial Information Sciences}, XLVIII-M-3-2023:81--85, 2023.

\bibitem{gu2021adversarialexamplessegmentationmodels}
Jindong Gu, Hengshuang Zhao, Volker Tresp, and Philip Torr.
\newblock {Adversarial Examples on Segmentation Models can be Easy to
  Transfer}, 2021.

\bibitem{dong2020benchmarking}
Yinpeng Dong, Qi-An Fu, Xiao Yang, Tianyu Pang, Hang Su, Zihao Xiao, and Jun
  Zhu.
\newblock {Benchmarking Adversarial Robustness on Image Classification}.
\newblock In {\em 2020 IEEE/CVF Conference on Computer Vision and Pattern
  Recognition (CVPR)}, pages 318--328, 2020.

\bibitem{liu2024task}
Chenying Liu, Conrad~M Albrecht, Yi~Wang, and Xiao~Xiang Zhu.
\newblock {Task Specific Pretraining with Noisy Labels for Remote Sensing Image
  Segmentation}.
\newblock In {\em IGARSS 2024-2024 IEEE International Geoscience and Remote
  Sensing Symposium}, pages 7040--7044. IEEE, 2024.

\bibitem{4382932}
Tamlin~M. Pavelsky and Laurence~C. Smith.
\newblock {RivWidth: A Software Tool for the Calculation of River Widths From
  Remotely Sensed Imagery}.
\newblock {\em IEEE Geoscience and Remote Sensing Letters}, 5(1):70--73, 2008.

\bibitem{8752013}
Xiao Yang, Tamlin~M. Pavelsky, George~H. Allen, and Gennadii Donchyts.
\newblock {RivWidthCloud: An Automated Google Earth Engine Algorithm for River
  Width Extraction From Remotely Sensed Imagery}.
\newblock {\em IEEE Geoscience and Remote Sensing Letters}, 17(2):217--221,
  2020.

\bibitem{klemenjak2012automatic}
Sascha Klemenjak, Bj\"{o}rn Waske, Silvia Valero, and Jocelyn Chanussot.
\newblock {Automatic Detection of Rivers in High-Resolution SAR Data}.
\newblock {\em IEEE Journal of Selected Topics in Applied Earth Observations
  and Remote Sensing}, 5(5):1364--1372, 2012.

\bibitem{Jinsong2021TGRSWaterSeg}
Jinsong Zhang, Mengdao Xing, Guang-Cai Sun, Jianlai Chen, Mengya Li, Yihua Hu,
  and Zheng Bao.
\newblock {Water Body Detection in High-Resolution SAR Images With Cascaded
  Fully-Convolutional Network and Variable Focal Loss}.
\newblock {\em IEEE Transactions on Geoscience and Remote Sensing},
  59(1):316--332, 2021.

\bibitem{long2015fully}
Jonathan Long, Evan Shelhamer, and Trevor Darrell.
\newblock {Fully Convolutional Networks for Semantic Segmentation}.
\newblock In {\em Proceedings of the IEEE Conference on Computer Vision and
  Pattern Recognition (CVPR)}, pages 3431--3440, 2015.

\bibitem{jegou2017one}
Simon J{\'e}gou, Michal Drozdzal, David Vazquez, Adriana Romero, and Yoshua
  Bengio.
\newblock {The One Hundred Layers Tiramisu: Fully Convolutional DenseNets for
  Semantic Segmentation}.
\newblock In {\em Proceedings of the IEEE Conference on Computer Vision and
  Pattern Recognition (CVPR) Workshops}, pages 11--19, 2017.

\bibitem{isola2017image}
Phillip Isola, Jun-Yan Zhu, Tinghui Zhou, and Alexei~A Efros.
\newblock {Image-to-Image Translation With Conditional Adversarial Networks}.
\newblock In {\em Proceedings of the IEEE Conference on Computer Vision and
  Pattern Recognition (CVPR)}, pages 1125--1134, 2017.

\bibitem{OBIDA2019101910}
Christopher~B. Obida, George~A. Blackburn, James~D. Whyatt, and Kirk~T. Semple.
\newblock {River Network Delineation from Sentinel-1 SAR Data}.
\newblock {\em International Journal of Applied Earth Observation and
  Geoinformation}, 83:101910, 2019.

\bibitem{qi2023dynamic}
Yaolei Qi, Yuting He, Xiaoming Qi, Yuan Zhang, and Guanyu Yang.
\newblock {Dynamic Snake Convolution Based on Topological Geometric Constraints
  for Tubular Structure Segmentation}.
\newblock In {\em Proceedings of the IEEE/CVF International Conference on
  Computer Vision (ICCV)}, pages 6070--6079, 2023.

\bibitem{verma2021deeprivwidth}
Ujjwal Verma, Arjun Chauhan, Manohara~Pai MM, and Radhika Pai.
\newblock {DeepRivWidth: Deep learning based semantic segmentation approach for
  river identification and width measurement in SAR images of Coastal
  Karnataka}.
\newblock {\em Computers \& Geosciences}, 154:104805, 2021.

\bibitem{chen2018encoder}
Liang-Chieh Chen, Yukun Zhu, George Papandreou, Florian Schroff, and Hartwig
  Adam.
\newblock {Encoder-Decoder with Atrous Separable Convolution for Semantic Image
  Segmentation}.
\newblock In {\em Proceedings of the European Conference on Computer Vision
  (ECCV)}, pages 801--818, 2018.

\bibitem{sen1floods119150760}
Derrick Bonafilia, Beth Tellman, Tyler Anderson, and Erica Issenberg.
\newblock {Sen1Floods11: A Georeferenced Dataset to Train and Test Deep
  Learning Flood Algorithms for Sentinel-1}.
\newblock In {\em 2020 IEEE/CVF Conference on Computer Vision and Pattern
  Recognition Workshops (CVPRW)}, pages 835--845, 2020.

\bibitem{warmedinger2023new}
Leena Warmedinger, Martin Huber, Mira Anand, Bernhard Lehner, Carolin Walper,
  Larissa Gorzawski, Michele Thieme, Birgit Wessel, and Achim Roth.
\newblock {The New Hydrographic Hydrosheds Database Derived from the Tandem-X
  Dem}.
\newblock In {\em Proceedings of 2023 IEEE International Geoscience and Remote
  Sensing Symposium (IGARSS 2023)}, pages 1485--1488. IEEE, 2023.

\bibitem{hijmans2001computer}
Robert~J Hijmans, Luigi Guarino, Mariana Cruz, and Edwin Rojas.
\newblock {Computer Tools for Spatial Analysis of Plant Genetic Resources Data:
  1. DIVA-GIS}.
\newblock {\em Plant genetic resources newsletter}, (127):15--19, 2001.

\bibitem{yamazaki2019merit}
Dai Yamazaki, Daiki Ikeshima, Jeison Sosa, Paul~D Bates, George~H Allen, and
  Tamlin~M Pavelsky.
\newblock {MERIT Hydro: A High-resolution Global Hydrography Map Based on
  Latest Topography Dataset}.
\newblock {\em Water Resources Research}, 55(6):5053--5073, 2019.

\bibitem{s1s210321672}
Marc Wieland, Florian Fichtner, Sandro Martinis, Sandro Groth, Christian
  Krullikowski, Simon Plank, and Mahdi Motagh.
\newblock {S1S2-Water: A Global Dataset for Semantic Segmentation of Water
  Bodies from Sentinel-1 and Sentinel-2 Satellite Images}.
\newblock {\em IEEE Journal of Selected Topics in Applied Earth Observations
  and Remote Sensing}, 17:1084--1099, 2024.

\bibitem{heller2018imperfect}
Nicholas Heller, Joshua Dean, and Nikolaos Papanikolopoulos.
\newblock {Imperfect Segmentation Labels: How much do they matter?}
\newblock In {\em Intravascular Imaging and Computer Assisted Stenting and
  Large-Scale Annotation of Biomedical Data and Expert Label Synthesis: 7th
  Joint International Workshop, CVII-STENT 2018 and Third International
  Workshop, LABELS 2018, Held in Conjunction with MICCAI 2018, Granada, Spain,
  September 16, 2018, Proceedings 3}, pages 112--120. Springer, 2018.

\bibitem{lad2023estimatinglabelqualityerrors}
Vedang Lad and Jonas Mueller.
\newblock {Estimating Label Quality and Errors in Semantic Segmentation Data
  via any Model}, 2023.

\bibitem{northcutt2021pervasive}
Curtis~G Northcutt, Anish Athalye, and Jonas Mueller.
\newblock {Pervasive Label Errors in Test Sets Destabilize Machine Learning
  Benchmarks}.
\newblock In {\em Thirty-fifth Conference on Neural Information Processing
  Systems Datasets and Benchmarks Track (Round 1)}, 2021.

\bibitem{ZHU2024128512}
Yanfei Zhu, Yaochi Zhao, Zhuhua Hu, Tan Luo, and Like He.
\newblock {A Review of Black-box Adversarial Attacks on Image Classification}.
\newblock {\em Neurocomputing}, 610:128512, 2024.

\bibitem{lin2021mlattackmodelsadversarial}
Jing Lin, Long Dang, Mohamed Rahouti, and Kaiqi Xiong.
\newblock {ML Attack Models: Adversarial Attacks and Data Poisoning Attacks},
  2021.

\bibitem{He_2024}
Zhixun He and Mukesh Singhal.
\newblock {VQUNet: Vector Quantization U-Net for Defending Adversarial Attacks
  by Regularizing Unwanted Noise}.
\newblock In {\em Proceedings of the 2024 7th International Conference on
  Machine Vision and Applications}, volume~17 of {\em ICMVA 2024}, page
  69–76. ACM, March 2024.

\bibitem{10.1145/3468920.3468933}
Cheng Gu, Wenxin Hu, and Wei Zheng.
\newblock {Domain Adaptation Using NDWI index for Water-Land Semantic
  Segmentation}.
\newblock In {\em Proceedings of the 2021 3rd International Conference on Big
  Data Engineering}, BDE '21, page 90–96, New York, NY, USA, 2021.
  Association for Computing Machinery.

\bibitem{9897807}
Pooya Tavallali, Vahid Behzadan, Azar Alizadeh, Aditya Ranganath, and Mukesh
  Singhal.
\newblock {Adversarial Label-Poisoning Attacks and Defense for General
  Multi-Class Models Based on Synthetic Reduced Nearest Neighbor}.
\newblock In {\em 2022 IEEE International Conference on Image Processing
  (ICIP)}, pages 3717--3722, 2022.

\bibitem{shafahi2018poison}
Ali Shafahi, W.~Ronny Huang, Mahyar Najibi, Octavian Suciu, Christoph Studer,
  Tudor Dumitras, and Tom Goldstein.
\newblock {Poison Frogs! Targeted Clean-Label Poisoning Attacks on Neural
  Networks}.
\newblock In S.~Bengio, H.~Wallach, H.~Larochelle, K.~Grauman, N.~Cesa-Bianchi,
  and R.~Garnett, editors, {\em Advances in Neural Information Processing
  Systems}, volume~31. Curran Associates, Inc., 2018.

\bibitem{freeman1993radiometric}
A~Freeman.
\newblock {Radiometric Calibration of SAR Image Data}.
\newblock {\em International Archives of Photogrammetry and Remote Sensing},
  29:212--212, 1993.

\bibitem{inproceedings}
David Small, Adrian Schubert, Betlem Rosich, and Erich Meier.
\newblock {Geometric and Radiometric Correction of ESA SAR Products}.
\newblock pages 1--6, 04 2007.

\bibitem{lee1994speckle}
Jong-Sen Lee, L~Jurkevich, Piet Dewaele, Patrick Wambacq, and Andr{\'e}
  Oosterlinck.
\newblock {Speckle Filtering of Synthetic Aperture Radar Images: A Review}.
\newblock {\em Remote Sensing Reviews}, 8(4):313--340, 1994.

\bibitem{cumming2005digital}
I.G. Cumming and F.H. Wong.
\newblock {\em {Digital Processing of Synthetic Aperture Radar Data: Algorithms
  and Implementation}}.
\newblock Number v.1 in Artech House remote sensing library. Artech House,
  2005.

\bibitem{chen2017rethinkingatrousconvolutionsemantic}
Liang-Chieh Chen, George Papandreou, Florian Schroff, and Hartwig Adam.
\newblock {Rethinking Atrous Convolution for Semantic Image Segmentation},
  2017.

\bibitem{xie2021segformer}
Enze Xie, Wenhai Wang, Zhiding Yu, Anima Anandkumar, Jose~M. Alvarez, and Ping
  Luo.
\newblock {SegFormer: Simple and Efficient Design for Semantic Segmentation
  with Transformers}.
\newblock In {\em Proceedings of the 35th International Conference on Neural
  Information Processing Systems}, NIPS '21, Red Hook, NY, USA, 2021. Curran
  Associates Inc.

\bibitem{cheng2022masked}
Bowen Cheng, Ishan Misra, Alexander~G Schwing, Alexander Kirillov, and Rohit
  Girdhar.
\newblock {Masked-Attention Mask Transformer for Universal Image Segmentation}.
\newblock In {\em Proceedings of the IEEE/CVF Conference on Computer Vision and
  Pattern Recognition}, pages 1290--1299, 2022.

\bibitem{liu2021swin}
Ze~Liu, Yutong Lin, Yue Cao, Han Hu, Yixuan Wei, Zheng Zhang, Stephen Lin, and
  Baining Guo.
\newblock {Swin Transformer: Hierarchical Vision Transformer using Shifted
  Windows}.
\newblock In {\em Proceedings of the IEEE/CVF International Conference on
  Computer Vision}, pages 10012--10022, 2021.

\bibitem{5206848}
Jia Deng, Wei Dong, Richard Socher, Li-Jia Li, Kai Li, and Li~Fei-Fei.
\newblock {ImageNet: A Large-scale Hierarchical Image Database}.
\newblock In {\em 2009 IEEE Conference on Computer Vision and Pattern
  Recognition}, pages 248--255, 2009.

\bibitem{cordts2016cityscapes}
Marius Cordts, Mohamed Omran, Sebastian Ramos, Timo Rehfeld, Markus Enzweiler,
  Rodrigo Benenson, Uwe Franke, Stefan Roth, and Bernt Schiele.
\newblock {The Cityscapes Dataset for Semantic Urban Scene Understanding}.
\newblock In {\em Proceedings of the IEEE conference on computer vision and
  pattern recognition}, pages 3213--3223, 2016.

\bibitem{CHE2025132771}
Lusheng Che, Shuangshuang Li, and Xianfeng Liu.
\newblock {Improved Surface Water Mapping Using Satellite Remote Sensing
  Imagery Based on Optimization of the Otsu Threshold and Effective Selection
  of Remote-Sensing Water Index}.
\newblock {\em Journal of Hydrology}, 654:132771, 2025.

\end{thebibliography}
\end{document}